\documentstyle[epsf,preprint,prl,aps]{revtex}
\begin{document}
\draft
\title{Theory of Chiral Order in Random Copolymers}
\author{J. V. Selinger}
\address{Center for Bio/Molecular Science and Engineering,
Naval Research Laboratory, Code 6900, \\
4555 Overlook Avenue, SW,
Washington, DC  20375}
\author{R. L. B. Selinger}
\address{Materials Science and Engineering Laboratory,
National Institute of Standards and Technology, \\
Gaithersburg, MD  20899}
\date{September 15, 1995}
\maketitle
\begin{abstract}
Recent experiments have found that polyisocyanates composed of a mixture
of opposite enantiomers follow a chiral ``majority rule:''  the chiral
order of the copolymer, measured by optical activity, is dominated by
whichever enantiomer is in the majority.  We explain this majority rule
theoretically by mapping the random copolymer onto the random-field
Ising model.  Using this model, we predict the chiral order as a function
of enantiomer concentration, in quantitative agreement with the
experiments, and show how the sharpness of the majority-rule curve can
be controlled.
\end{abstract}
\narrowtext

Cooperative chiral order plays a vital role in the self-assembly of
ordered supramolecular structures in liquid crystals~\cite{lc}, organic
thin films~\cite{film1,film3,film4,film5}, and lipid
membranes~\cite{tubule0,tubule1,tubule2,tubule3,tubule4,tubule5,ripple}.
One particularly simple and well-controlled example of cooperative chiral
order is in random copolymers.  Recent experiments have found that
polyisocyanates formed from a mixture of opposite enantiomers follow a
chiral ``majority rule''~\cite{majority}.  The chiral order of the
copolymer, measured by optical activity, responds sharply to slight
differences in the concentrations of the enantiomers, and is dominated by
whichever enantiomer is in the majority.  In this paper, we show that
the majority rule can be understood through a mapping of the random
copolymer onto the random-field Ising model~\cite{rfim1,rfim2,rfim3}.
Using this model, we predict the chiral order as a function of enantiomer
concentration, in quantitative agreement with the experiments, and
show that the sharpness of the majority-rule curve is determined by two
energy scales associated with the chiral packing of monomers.

\begin{figure}
\epsfbox{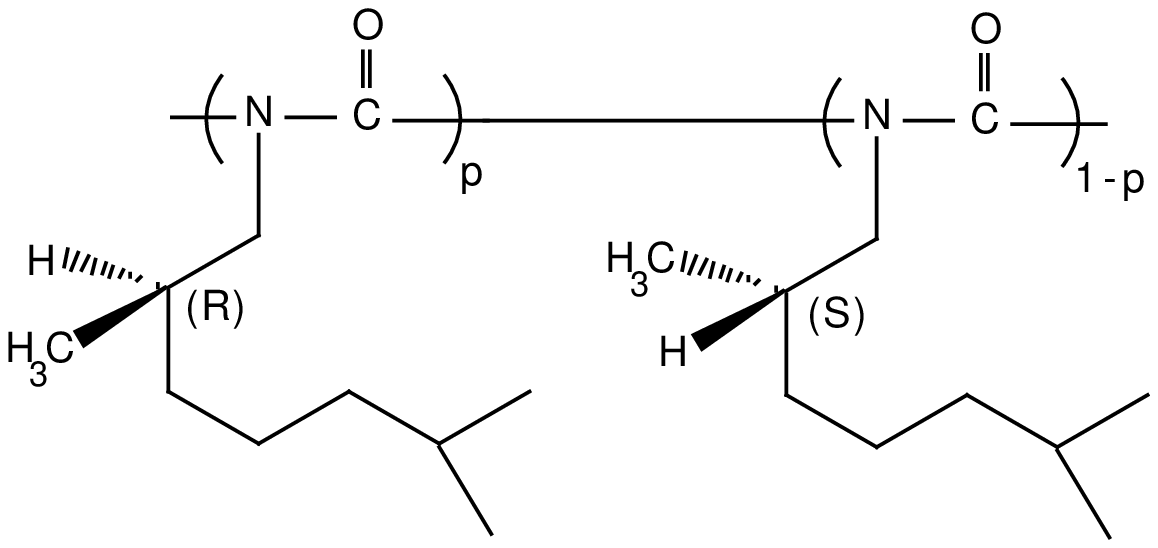}
\caption{Molecular structure of the polyisocyanate with enantiomeric
pendant groups derived from citronellic acid, which was studied in
Ref.~\protect\cite{majority}.}
\label{fig1}
\end{figure}

\begin{figure}
\epsfbox{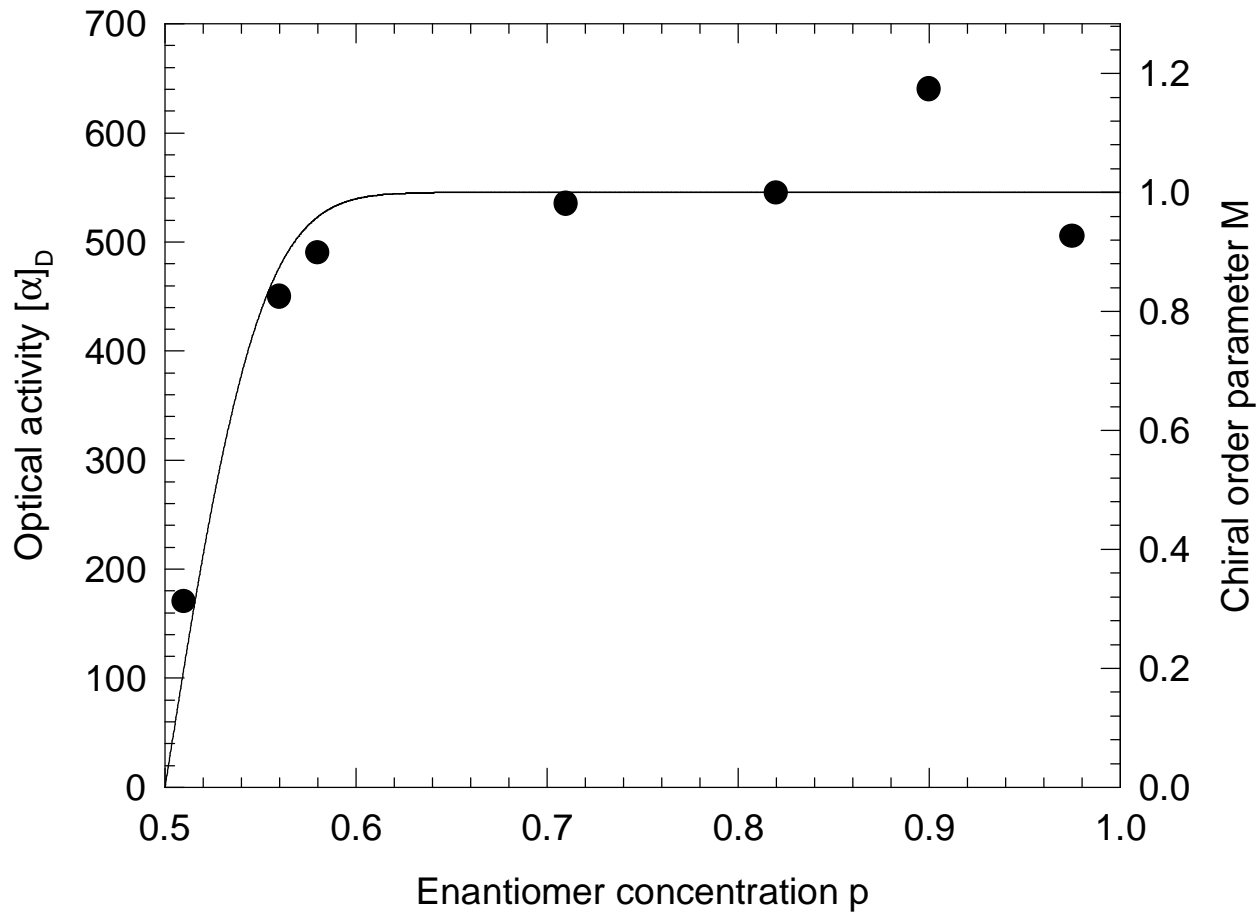}
\caption{Symbols:  Optical activity $[\alpha]_D$ of the random copolymer
at the sodium D-line as a function of enantiomer concentration $p$, from
Ref.~\protect\cite{majority}.  Solid line:  Theoretical prediction for
the chiral order parameter $M$ from Eq.~(\protect\ref{erfeq}), with the
domain size $L=164$.  The prediction agrees with the data with no
adjustable parameters, other than the relative scale of the vertical
axes.}
\label{fig2}
\end{figure}

In a series of experiments, Green {\it et al.}\ have investigated chiral
order in polyisocyanates~\cite{picreview}.  This polymer consists of a
carbon-nitrogen backbone with a pendant group attached to each monomer,
as shown in Fig.~\ref{fig1}.  Although the backbone is nonchiral, steric
constraints force the molecule to polymerize in a helical structure.  If
the pendant group is also nonchiral, the helix is randomly right- or
left-handed.  A long chain then consists of domains of fixed helicity,
separated by occasional helix reversals.  On average, there are equal
right- and left-handed domains, leading to zero net optical activity.
However, if the pendant group is chiral, there is a preference for one
sense of the helix, which leads to a net optical activity.  Because of
the cooperative interaction between the monomers in a domain, even a very
small chiral influence leads to a large optical
activity~\cite{deuterium}.  Most recently, Green {\it et al.}\ have
synthesized random copolymers with a mixture of right- and left-handed
enantiomeric pendant groups, with concentrations $p$ and $1-p$,
respectively~\cite{majority}.  The resulting optical activity, shown in
Fig.~\ref{fig2}, has a surprisingly sharp dependence on $p$.  A 56/44
mixture of enantiomers has almost the same optical activity as a pure
100/0 homopolymer, and even a 51/49 mixture has a third of that optical
activity.

To explain this cooperative chiral order theoretically, we map the random
copolymer onto the one-dimensional random-field Ising model, a standard
model in the theory of random magnetic systems~\cite{rfim1,rfim2,rfim3}.
Although related models have been applied to other polymer
systems~\cite{copoly1,copoly2,copoly3,copoly4}, our theory gives a new,
direct correspondence between the Ising order parameter and the optical
activity.  This correspondence provides a novel experimental test of
predictions for the random-field Ising model.  Let the Ising spin
$\sigma_i=\pm 1$ represent the sense of the polymer helix at the monomer
$i$.  The energy of a polymer can then be written as
\begin{equation}
H=-J\sum_{i=1}^{N-1}\sigma_i\sigma_{i+1}-\sum_{i=1}^{N}h_i\sigma_i.
\end{equation}
Here, the random field $h_i$ specifies the enantiomeric identity of the
pendant group on monomer $i$:  $h_i=+h$ if it is right-handed (with
probability $p$) and $h_i=-h$ if it is left-handed (with probability
$1-p$).  This field is a {\it quenched\/} random variable; it is fixed by
the polymerization of each individual chain.  The parameter $2h$ is the
energy cost of a right-handed monomer in a left-handed helix, or vice
versa.  Molecular modeling gives $2h \approx 1.7$~kJ/mol (0.4~kcal/mol)
for the pendant group used in these experiments\cite{molecmodel}.  The
parameter $2J$ is the energy cost of a helix reversal.  Fits of the
optical activity of pure homopolymers as a function of temperature $T$
give $2J \approx 17$~kJ/mol (4~kcal/mol)~\cite{deuterium}.  Note that
$2J$ is much greater than $k_B T \approx 2.5$~kJ/mol (0.6~kcal/mol), but
$2h < k_B T$.  The degree of polymerization $N$ ranges from 350 to 5800
in the experiments~\cite{majority}.  The magnetization of the Ising model,
\begin{equation}
M=\left\langle\frac{1}{N}\sum_{i=1}^{N}\sigma_i\right\rangle,
\end{equation}
corresponds to the chiral order parameter that is measured by optical
activity.  To predict the optical activity of the random copolymer as a
function of enantiomer concentration, we must calculate $M$ as a function
of $p$.

To calculate the order parameter $M$, we note that each chain consists of
domains of uniform helicity $\sigma_i$.  As an approximation, suppose
that each domain has length $L$, which is to be determined.  Each domain
responds to the total chiral field $h_{\rm tot}=\sum h_i$ of the monomers
in it.  Because the domain is uniform, the response is
$M(h_{\rm tot})=\tanh(h_{\rm tot}/k_B T)$, equivalent to a single spin in
a magnetic field.  Averaging over the probability distribution
$P(h_{\rm tot})$, we obtain
\begin{equation}
M=\int_{-\infty}^{\infty}dh_{\rm tot}P(h_{\rm tot})
\tanh\left(\frac{h_{\rm tot}}{k_B T}\right).
\end{equation}
The probability distribution $P(h_{\rm tot})$ is a binomial distribution.
For large domains, it can be approximated by a Gaussian with mean
${2hL(p-\frac{1}{2})}$ and standard deviation ${2h[Lp(1-p)]^{1/2}}$.  For
$p\approx\frac{1}{2}$, the standard deviation becomes $hL^{1/2}$.
Furthermore, if the width of the Gaussian is much greater than the width
of the tanh, $hL^{1/2}\gg k_B T$, then the tanh can be approximated by a
step function.  The expression for $M$ then becomes
\begin{equation}
M\approx{\rm erf}\left[\left(2L\right)^{1/2}
\left(p-\frac{1}{2}\right)\right].
\label{erfeq}
\end{equation}

We must now estimate the domain size $L$.  The domain size is determined
by (a)~the distance $L_{\rm rf}$ between helix reversals that are induced
by the random field, (b)~the distance $L_{\rm th}$ between helix
reversals that are induced by thermal fluctuations, and (c)~the chain
length $N$.  Because each of these effects contributes to the density
$1/L$ of domain boundaries, we expect
\begin{equation}
\frac{1}{L}\approx\frac{1}{L_{\rm rf}}+\frac{1}{L_{\rm th}}+\frac{1}{N}.
\end{equation}
For $p\approx\frac{1}{2}$, the random-field domain size $L_{\rm rf}$
can be estimated using a variation of the Imry-Ma argument for the
random-field Ising model~\cite{rfim1}.  A domain forms when the field
energy $hL^{1/2}$ grows to equal the boundary energy $J$.  By equating
these two energies, we obtain the random-field domain size
\begin{equation}
L_{\rm rf}\approx\left(\frac{J}{h}\right)^2 .
\label{imryma}
\end{equation}
With the values $2J \approx 17$~kJ/mol (4~kcal/mol) and
$2h \approx 1.7$~kJ/mol (0.4~kcal/mol) appropriate for the experimental
system, the random-field domain size becomes
$L_{\rm rf} \approx 100$~monomers.  By comparison, the thermal domain
size is $L_{\rm th}=e^{2J/k_B T} \approx 800$, and the chain length is
$N \approx 350$--$5800$.  Because $L_{\rm rf}$ is much less than
$L_{\rm th}$ and $N$, we obtain $L\approx L_{\rm rf}$; i.~e., the domain
size is limited by random-field effects.

\begin{figure}
\epsfbox{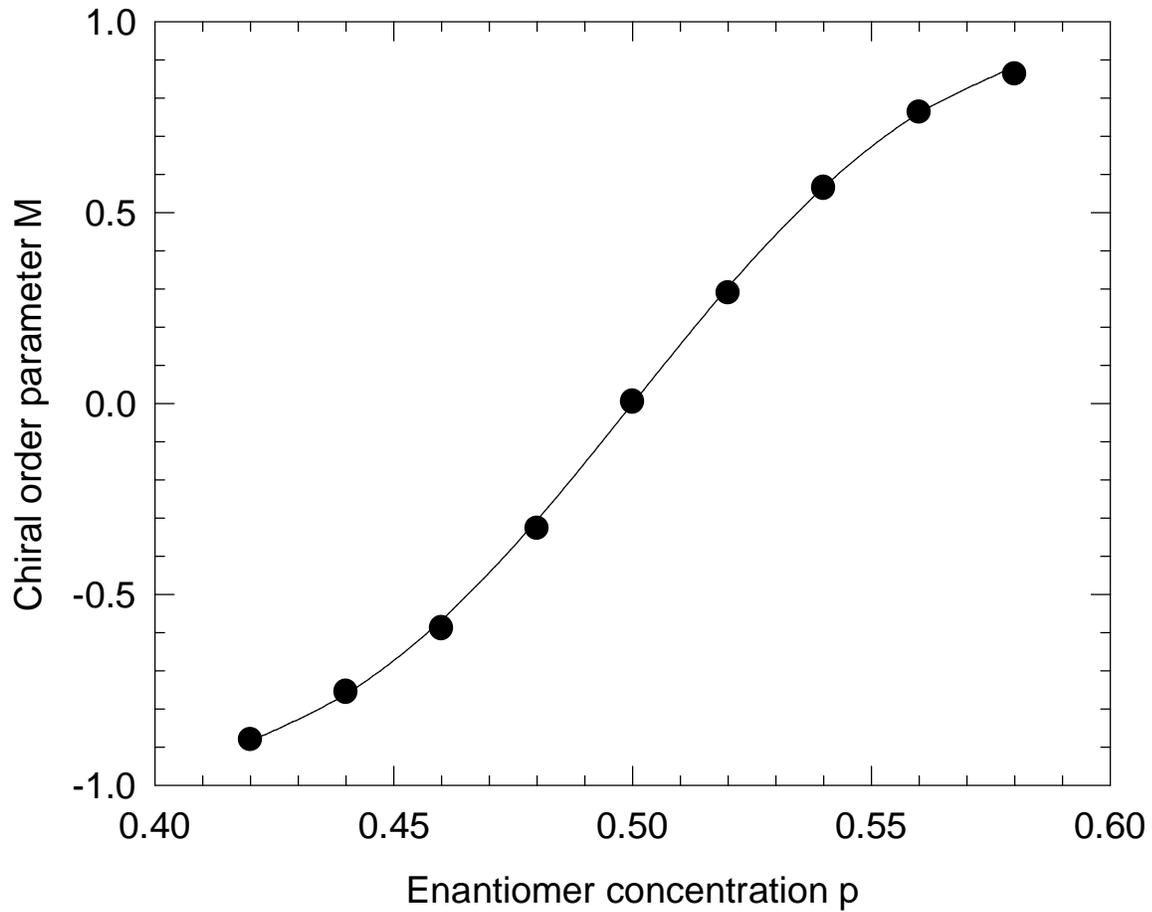}
\caption{Numerical simulation of the chiral order parameter $M$ as a
function of enantiomer concentration $p$ for chain length $N=230$.  The
solid line shows a fit to the prediction of Eq.~(\protect\ref{erfeq}),
which gives the domain size $L(230)=96$.}
\label{fig3}
\end{figure}

\begin{figure}
\epsfbox{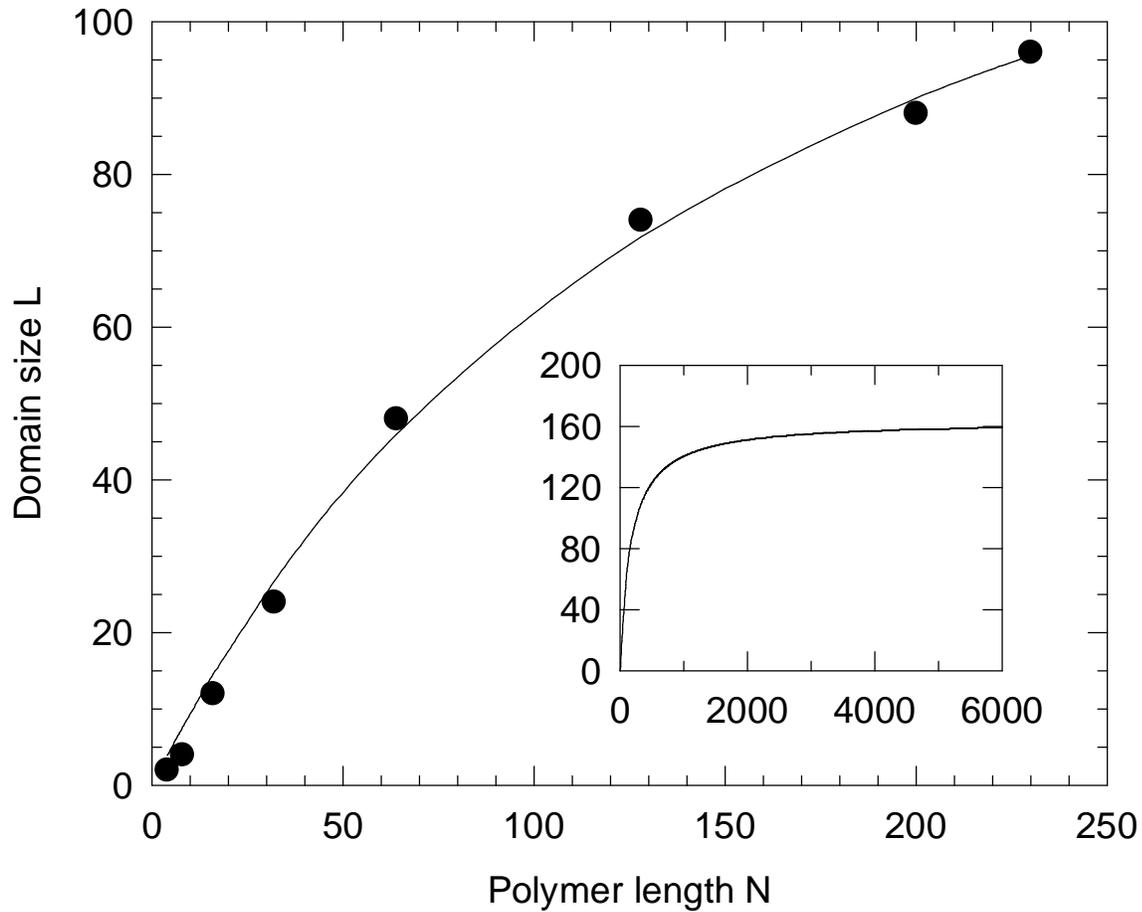}
\caption{The domain size $L(N)$ from numerical simulations for several
values of the chain length $N$.  The solid line shows a fit to the
extrapolation form $1/L(N)=1/L_{\rm max}+1/N$, or
$L(N)=L_{\rm max} N/(L_{\rm max}+N)$, which gives $L_{\rm max}=164$.
The inset shows the extrapolation up to $N=6000$.}
\label{fig4}
\end{figure}

To test this approximate calculation explicitly, and to obtain a more
precise value of the domain size, we performed numerical simulations of
the random-field Ising model.  In these simulations, we used a series of
chain lengths from 4 to 230, and used the values of $J$ and $h$
appropriate for the experiments.  For each chain length, we constructed
an explicit realization of the random field, then calculated the
partition function and order parameter using transfer-matrix techniques.
We then averaged the order parameter over at least 1000 realizations of
the random field.  Figure~\ref{fig3} shows $M$ as a function of $p$ for
chain length $N=230$.  These results can be fit very well to
Eq.~(\ref{erfeq}), with the domain size $L(230)=96$.  The results for
other values of $N$ can be fit equally well.  Figure~\ref{fig4} shows the
fitted domain size $L(N)$ as a function of chain length $N$.  These
results can be extrapolated using $1/L(N)=1/L_{\rm max}+1/N$, with
$L_{\rm max}=164$.  This random-field domain size agrees well with
Eq.~(\ref{imryma}), especially considering that the Imry-Ma argument is
only a scaling argument.  Thus, the chiral order parameter should indeed
be given by Eq.~(\ref{erfeq}), with the extrapolated domain size $L=164$.

To compare our theory with the experiment, we plot our prediction for the
chiral order parameter $M$ on top of the experimental data for the
optical activity in Fig.~\ref{fig2}.  The prediction agrees very well
with the data.  In particular, $M$ saturates at
$(p-\frac{1}{2})\approx 0.06$, a 56/44 composition, in agreement with the
data.  This saturation point is a direct measure of $1/(2L)^{1/2}$, which
is controlled by the ratio of the two energy scales $J$ and $h$.  We
emphasize that the theory matches the experimental data with
{\it no adjustable parameters,} other than the relative scale of the
optical-activity axis and the order-parameter axis.  That relative scale
is the optical activity of a pure 100/0 homopolymer.

We can make two remarks about these results.  First, both the
random-field domain size $L_{\rm rf}$ and the thermal domain size
$L_{\rm th}$ depend on $p$.  The Imry-Ma argument above applies only to
the regime where $p\approx\frac{1}{2}$.  This is the appropriate regime
for understanding the experiments, because all the significant variation
in the chiral order parameter $M$ occurs around $p\approx\frac{1}{2}$.
Outside that regime, $M$ saturates at $\pm 1$, and it is not sensitive
to $L$.  Second, the fact that $L_{\rm rf} \ll L_{\rm th}$ shows that
quenched disorder is much more significant than thermal disorder in
this copolymer system.  Thus, the system is effectively in the
low-temperature limit.  If the temperature $T$ were increased so that
$L_{\rm th} \lesssim L_{\rm rf}$, then thermal disorder (i.~e. entropy)
would become more significant.  However, that would require a
temperature of $T \gtrsim 400$~K, which is unrealistic because the
system would degrade chemically.

Using our theory, we can make two predictions for future experiments.
First, the experiment could be repeated using different degrees of
polymerization.  For short chains, the domain size $L$ is limited by the
chain length $N$, particularly for $N\lesssim 200$, as shown in
Fig.~\ref{fig4}.  Thus, shorter chains should give a {\it broader\/}
version of the majority-rule curve.  By contrast, longer chains should
{\it not\/} give a sharper majority-rule curve, because the chains are
already in the regime where $L$ is approximately independent of $N$.
Second, the experiment could be repeated using pendant groups that are
``less chiral,'' i.~e. polyisocyanates with a lower energy cost for a
right-handed monomer in a left-handed helix.  A lower value of the chiral
field $h$ should give a larger value of the domain size
$L\approx(J/h)^2$, and hence a {\it sharper\/} majority-rule curve.
(This prediction applies as long as $L_{\rm rf}\ll L_{\rm th}$, or
$2h\gg 2Je^{-J/k_B T}\approx 0.6$~kJ/mol [0.14~kcal/mol].  Thus, $h$ can
be reduced by a factor of 3 from its value in the current experiments,
and the majority-rule curve can become 3 times sharper.  Beyond that
point, the sharpness will be limited by the thermal domain size.)  This
second prediction might seem counter-intuitive, because one might expect
a smaller chiral field to give a smaller effect.  However, this
prediction is reasonable, considering that the majority-rule curve is
limited by the number of monomers that cooperate inside a single domain.
If the local chiral field is reduced, then each monomer is more likely to
have the same helicity as its neighbors, independent of the local chiral
field, and hence the cooperativity increases.

Finally, we note that the sharp majority-rule curve in polyisocyanates can
be exploited in an optical switch~\cite{switch}.  If a mixture of
enantiomers is exposed to a pulse of circularly polarized light, one
enantiomer is preferentially excited into a higher-energy state.  That
state can decay into either chiral form.  Thus, a light pulse depletes the
preferentially excited enantiomer and changes the enantiomer concentration
$p$, which changes the optical activity.  In other systems, this approach
has been limited by the fact that light pulses induce only a slight change
in $p$, and hence only a slight change in optical activity.  However, in
polyisocyanates, a slight change in $p$ close to $p=\frac{1}{2}$ is
sufficient to induce a very significant change in optical activity.
Indeed, this polymer has almost a binary response to changes in $p$, which
is needed for an optical switch.  Our theory shows how to optimize the
majority-rule curve for use in an optical switch.

In conclusion, we have shown that the cooperative chiral order in
polyisocyanates can be understood through the random-field Ising model.
The energy scales $J$ and $h$, which arise from the chiral packing of the
monomers, give the random-field domain size $L_{\rm rf}$, which indicates
how many monomers are correlated in a single domain.  This domain size
determines the sharpness of the majority-rule curve.  Our theory agrees
well with the current experiments, and it shows how future experiments
can control the chiral order.

We thank M. M. Green and J. M. Schnur for many helpful discussions.

\end{document}